# The use of epistemic distancing to create a safe space to sensemake in introductory physics tutorials

Luke D. Conlin[1]

[1]*Stanford University, Graduate School of Education,*
*150 Serra Mall, Stanford, CA, 94305*

**Abstract.** In active engagement physics classrooms, students get opportunities to make sense of physics together through discussion. They do not always take up these opportunities, in part because of the risk of sharing their ideas and having them rejected by their classmates or the instructors. In this case study, I analyze videotaped discourse of a tutorial group's early discussions to investigate how students manage these risks in creating a safe space to sensemake. I find that the students and instructors alike rely on a common discursive resource – epistemic distancing – to share their ideas while protecting themselves affectively if others disagree. Epistemic distancing includes hedging, joking, deferring, and other discourse moves used to soften one's stance in conversation. I use video analysis to illustrate the effects of these moves on tutorial groups' sensemaking discussions, and discuss implications for instructors wishing to encourage sensemaking discussions in their physics classrooms.

**PACS:** 01.40.-d, 01.40.Fk, 01.40.Ha, 01.40.G-, 01.50.ht

## I. INTRODUCTION

Active engagement classrooms give students opportunities to learn physics by discussing their ideas with their classmates. For instance, in the *Tutorials in Physics Sensemaking* [1] students work together in groups of four to make sense of conceptual and epistemological issues in physics through worksheet-guided discussions. Discussing their ideas involves risk—including the risk of sharing ones' ideas and possibly having them rejected by their peers or by the instructor. If students are to engage in sensemaking discussions, they must find ways to manage these risks.

To understand how tutorial groups first created a safe space to sensemake within their discussions, I analyzed three student groups' first discussions of the semester. Here I present a case study of one of the groups to illustrate the construct of epistemic distancing and its critical role in the groups' constructing a safe space to discuss their physics ideas. I find that the students and instructors use hedging, joking, quoting and other discourse moves I classify as *epistemic distancing* to soften their stance and make room for sensemaking.

This paper first characterizes discourse moves that involve epistemic distancing. Then it presents a case study of two early discussions of one group to illustrate how students' and instructors' use of epistemic distancing makes space for the group to collaboratively make sense of physics.

## II. EPISTEMIC DISTANCING

Discourse analysts have highlighted how people index their stance in conversation to manage conflicts {2].

Hedging, by using phrases such as "I think," or "I guess" is one way of softening one's stance, as is quoting another person or deferring to others [2, 3]. Students use these distancing moves to manage conflict by expressing a contrary position without escalating the conflict, for example on collaborative writing tasks [2].

In active engagement classrooms in which students discuss their ideas, conflict is even more likely to occur, in part because students are likely to have different ideas about the same phenomenon. But little research has explored how students manage conflict in face-to-face interactions in science classrooms. In such environments, words are not the only way to downgrade epistemic stance. Claims can be softened by phrasing them as a question, or by simply using a rising intonation that suggests uncertainty [4]. Epistemic stance can be downgraded based on body positioning as well [5].

Any discourse move by which speakers soften their stance, including hedging, joking, and deferring, I characterize as *epistemic distancing*. Epistemic distancing is "epistemic" in that it concerns the speaker's commitment to the truth of what they are saying. It is "distancing" in that it creates distance between the speaker and what they are saying [6]. This distance protects the speaker's affect in the event that the idea gets evaluated negatively, thereby reducing ego threat.

In tutorial, students can distance themselves from their ideas about physics or from their ideas about how to go about learning physics together. In constructing a safe space to make sense of physics together, students must establish a norm of contributing their own ideas, and a norm of collaboratively evaluating those ideas. This mirrors the essential tension of science between generating hypotheses and critically evaluating those hypotheses [7, 8].

Epistemic distancing can help students evaluate the ideas, rather than the person coming up with them. This protects students from negative evaluation of their ideas, so that they are not discouraged from contributing more ideas to the discussion.

This is not to say that more epistemic distancing is necessarily better. Students who distance themselves too much from their ideas about physics or how to learn physics can come across as so dismissive of the activity that they discourage further contributions.

I find that the tutorial students use epistemic distancing in ways that are consequential for making a safe space to sensemake. I will detail how one group first constructed a safe space to make sense of physics, illustrating the critical role that epistemic distancing played in this process.

## III. METHODS

This paper presents a case study in which video records of students' discourse in tutorials were analyzed [9] to gain insight on the communicative processes by which groups enter into collaborative ensemaking discussions. Episodes were selected in which the group behaviorally oriented toward "having a discussion" [10], and in which that discussion involved making sense of physical mechanisms.
Clip selection.
The full analysis focuses on three tutorial groups that provided a range of levels of engagement in the tutorials 11]. Using the methods described in Scherr & Hammer [10], I identified all of the times each group was engaged in discussion throughout the semester. To analyze how groups were able to initially construct a safe space to sensemake, I selected each group's very first discussion, as well as their first discussions containing significant evidence of mechanistic reasoning [12]. I focus the analysis here on just one of the groups to provide insight on the discursive processes by which they constructed a safe space to sensemake.

### A. Instructional context

In tutorial, students meet weekly for 50 minutes for a worksheet-guided inquiry into select topics in physics. The worksheets focus on developing conceptual understandings in physics, as well as epistemological understandings about how to approach learning physics.

## IV. DATA & ANALYSIS

I focus the analysis on two early discussions of one group, consisting of four students (pseudonyms: Alan, Brandi, Chrissie, and Daria). Both discussions take place during the first tutorial session of the semester, which explores conceptual and epistemological issues related to learning about motion graphs. In this tutorial, the students use a motion detector to make real-time graphs of their motion, which is displayed on a computer screen.

### A. Episode 1 – "Whatever. Next!"

The very first question of the first tutorial asks students to consider the benefits of discussing their mistakes in physics, and to write down their thoughts. It then asks them to discuss their answers, making a note of any differences. The Sparrows, who had been hunched over their respective worksheets, reading and writing, mark their transition to a group discussion by looking at each other and laughing. Then they very briefly discuss the first question:

CHRISSIE: *((laughs))*
DARIA: So…okay…we talked about how you can learn from your mistakes pretty much yeah.
ALAN: Yeah I think everyone said 'learning from your mistakes,' right?
DARIA: Yeah.
BRANDI: Right.
CHRISSIE: *((laughs))*
DARIA: Pretty much okay.
ALAN: Whatever. Next!

This is a case of too much epistemic distancing. Chrissie's laughter implies a lack of seriousness. Daria's "we" and "pretty much" distance themselves from their contribution of how to proceed. Overall they epistemically distance themselves in a way that they have the effect of the group avoiding discussing their ideas.

The group continues in this dismissive approach to the tutorial questions until later on in the first tutorial, when a teaching assistant (TA) overhears a good question and uses it as an opportunity to get them discussing their ideas.

### B. Episode 2 – "Just getting started up"

Later in the same tutorial session, the group is working on a question about motion graphs. The worksheet asked them to make a prediction of a distance vs. time graph of a person walking slowly and steadily away from a motion detector. They had all predicted that it would make a straight line with positive slope (did they discuss this?) Then the worksheet asked them to pick a person to physically make the graph.

Alan takes on the role of walker. He starts half a meter away from the detector, holding a book up as a target, and then walks slowly back. As he is returning to the table, he notices two unexpected "jumps" in the graph:

ALAN: Wh- what are those two jumps?
DARIA: ((Laughing)) Heh- I don't know.
ALAN: Whatever. ((Sits down))
CHRISSIE: Okay, ((reading out loud and trailing off)) "Sketch the result"
TA JOEY: So wait a second, that's a good question. What are those two jumps?

The group is continuing in their dismissive approach to the tutorial. Even though Alan asked a question to the group about two apparent deviations from the graph, the group does not discuss it, because of too much epistemic distancing. Daria laughs and says, "I don't know." Alan says "Whatever" as he sits down and starts to move on. Chrissie starts to read the next tutorial question, deferring to the worksheet. All of these moves increase the epistemic distancing so much that there is no space to have a discussion.

TA Joey had overheard Alan's question and uses it to try to get them discussing. He revoices (or just poses the question to the group), but again the only response is "I don't know". TA Joey modifies the question by adding increasing levels of epistemic distancing. He rephrases the question, "So, what do you think happened there, do you have any idea?" This version of the question incorporates considerable hedging, both by adding "what do you think" (from "what happened there" to "what do you think happened there") and asking for "any idea", which opens up space for someone to offer an idea even if they do not know for sure. As he asks this question with hedging, he also kneels down to eye level with the students. [Fig 1].

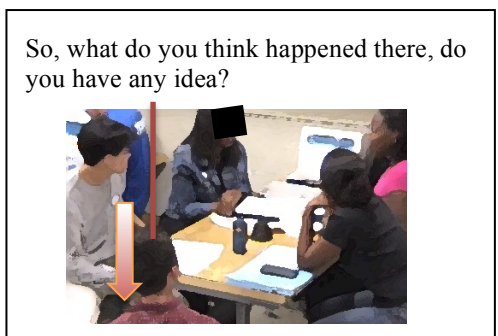

So, what do you think happened there, do you have any idea?

**FIG 1.** The TA kneels down as he rephrases the question with more epistemic distancing.

At this point, the students start to offer ideas about what may have happened:

| DARIA: | Something wrong must have happened. |
| ALAN: | I dunno, maybe…this (detector) was weird? |
| DARIA: | ((laughs)) |
| TA JOEY: | Maybe it was weird. What do you mean by- 'Weird' could mean a lot of things. |
| DARIA: | Maybe it's just getting started up or something. |
| TA JOEY: | It was getting started up, so like if we did it again, like now it's warmed up almost |
| DARIA: | Mayyybe |
| ALAN: | Maybe... |
| CHRISSIE: | I think we should do a second trial, to see |

Now that TA Joey has asked the question with more epistemic distancing, the group is willing to start tentatively coming up with and sharing some ideas. Daria starts off with a vague idea that "something wrong" happened. Alan further specifies that maybe the detector that was "weird". Although Alan may be half-joking (Daria laughs), TA Joey takes his idea seriously by echoing it back to him and asking him to be more specific. Daria offers an idea, "Maybe it's just getting started up or something" which she hedges by leading with "maybe" and ending with "or something". TA Joey also takes her idea seriously by echoing it once again, and pushes it further to infer a prediction: "so like if we did it again…" Chrissie then suggests they do another trial.

At this time, TA Joey tells the group that "this is the sort of thing we want you to investigate…this MOSTLY fits with your prediction, but there's some discrepancies…" and suggests they try it again. TA Joey leaves, and the group does another trial, with Alan walking slowly away. The group watches the screen carefully and as a straight line is drawn, this time with no "jumps". Daria exclaims, "THERE you goooo!" and Chrissie adds, "Ahhhh, okay!"

The group laughs and starts to move on to fill in the worksheet, but Chrissie keeps the discussion going by offering another idea of what might have happened:

| CHRISSIE: | So maybe you weren't walkin' at a steady pace at one point |
| ALAN: | Probably, I probably moved the book or something like that |
| DARIA: | Did you? Yeah maybe. |
| ALAN: | Yeah. |
| DARIA: | Wait did you do something different the first time? |
| ALAN: | No. |
| DARIA: | Like, while you were walking back? |
| ALAN: | I was- I probly…I donno either= |
| BRANDI: | Sometimes you do things subconsciously |
| ALAN: | =moved the book down or, you know, yeah. |

Even though TA Joey is no longer there, the group does not continue their dismissive approach to tutorial. They conduct another trial, and are happy when it yields better results. But they are not satisfied with the better graph; they continue his push to seek to understand what may have caused the discrepancy. To understand the different results, they discuss their ideas about the mechanisms behind the phenomena they are observing. In explaining the mysterious jumps, they are considering whether Alan had walked at a steady pace, or if he had inadvertently lowered the book he was holding as a target for the detector, and how he might not be aware of these subtle differences in his movement from one trial to the next.

Although there is not sufficient space to expand on this group's dynamics for the rest of the semester, in summary the group continues to have frequent discussions throughout the semester in which they make sense of the mechanisms brought up in tutorial.

## V.   CONCLUSIONS

In active engagement physics classrooms, students are given opportunities to share their ideas as they work collaboratively to understand phenomena.   But to share ones' ideas about the phenomena involves risk.   Little is understood about how students and teachers alike manage this risk in creating a safe space to sensemake.   I have shown a case study of how one group came to incorporate making sense of mechanisms into their discussions.   It took their instructor overhearing a good question, and using it as an opportunity to make sense of the phenomenon.   It was his introduction of epistemic distancing that made space for the students to offer their ideas of what was making the "jumps" in the graph.   When they introduced their ideas, they did so tentatively, with hedging and laughter, but the instructor took these ideas seriously, and pushed them to explore the mechanism behind the phenomena. Once the instructor left, they continued to discuss the mechanism behind the phenomena.   Their use of epistemic distancing, in response to TA Joey's, helped them switch out of their dismissive approach to tutorial and into one of deep inquiry.

## ACKNOWLEDGEMENTS

This work was funded by the National Science Foundation (Grant #0440113).   It builds on work by Rachel Scherr, who first noticed the phenomenon of tutorial students contributing ideas half-jokingly.